\def\exp{\hbox{exp}}
\def\ln{\ell{{\hbox {n}}}}
\documentstyle[a4,12pt]{article}

\begin{document}
\pagenumbering{arabic}

\begin{titlepage} \vspace{0.2in} \begin{flushright}
MITH-97/11 \\ \end{flushright} \vspace*{1.5cm}
\begin{center} {\LARGE \bf  On A Possible Ground State for Quantum 
Gravity$^{(*)}$
\\} \vspace*{0.8cm}
{\bf G.~Preparata and S.-S.~Xue}\\ \vspace*{1cm}
Dipartimento di Fisica dell'Universit\`a di Milano
\\INFN - Sezione di Milano, Via Celoria 16, Milano, Italia
\\ \vspace*{1.8cm}

{\bf   Abstract  \\ } \end{center} \indent

A variational calculation is presented of the ADM-energy of the 
quantized gravitational field around a wormhole solution of the 
classical Einstein's equations. One finds the energy of such state to 
be in general lower than the perturbative ground state, in which the 
quantized gravity field fluctuates around flat (Euclidean) space-time.
As a result the strong indication emerges that a gas (or a lattice) of
wormholes of Planck mass and average distance $l_p$, the Planck length,
may be a good approximation of the Ground State of Quantum Gravity, 
some implications of which are reviewed.

\vfill \begin{flushleft}  November, 1997   \vspace*{3cm} \\
\noindent{\rule[-.3cm]{5cm}{.02cm}} \\
\vspace*{0.2cm} \hspace*{0.5cm} $^{(*)}$
Presented at the 8th Marcel Grossmann Meeting, Jerusalem, 21-28, June 
1997\end{flushleft} \end{titlepage}

\section{The background: the ground state of QCD}

The problem of determining the Ground State (GS) --the vacuum -- of a 
Quantum Field Theory (QFT) is of central importance for a correct 
assessment and understanding of the dynamics of its observable excited 
states. Unfortunately after almost 70 years of QFT the solution of 
this very difficult problem in the most general context still escapes 
us. On the other hand, it must be admitted, most theoretical 
physicists believe that a good approximation, at least for small 
coupling (such as in QED, and Asymptotically Free QCD), is provided by 
the Perturbative Ground State (PGS), where the quantized field 
oscillators perform their {\it incoherent} zero-point fluctuations. 
Indeed the remarkable successes of QED and the much more problematic ones 
of Perturbative QCD (PQCD) have convinced the vast majority of 
physicists that for practical purposes, at least within the Standard 
Model (SM), the problem of the vacuum is essentially solved.

Between 1984 and 1986 the present author has developed a research 
program\cite{prep1,gp} aimed at ascertaining whether the PGS of QCD were 
really the good approximation to the vacuum of QCD that most 
believed, especially in view of the fact that the PGS leaves the 
fundamental question of colour confinement completely unanswered. Based 
on a gauge-invariant variational approach, this program led to the 
disquieting result that the PGS of QCD is ``essentially unstable'', 
i.e., contrary to what is generally believed, there exists {\it no}
space-time scale below which the PGS is a good approximation to {\it 
real} GS.

The basic idea behind such rather difficult calculation (which has 
been checked and confirmed by computer Monte Carlo 
simulations\cite{cosm}) was to probe the stability of the PGS by 
analyzing the effective quantum potential -- the energy density -- of 
the QFT in a classical background field, solution of the classical 
Yang-Mills equations of QCD. Following a suggestion by 
G.~Savvidy\cite{sav} a constant classical chromomagnetic field $B$ was 
chosen as background, and for $SU(2)$ the effective potential was 
found to differ from the patently faulty\cite{sav} result of the one-loop 
calculation by Savvidy\footnote{ in fact, as pointed out by Nielsen and 
Olesen in 1978\cite{no}, Savvidy's calculation had neglected, without 
justification, a set of field modes -- the ``unstable'' modes'' --
for which the frequencies for small oscillations are 
imaginary.}($\Lambda$ is the ultraviolet cut-off),
\begin{equation}
E(B) = {B^2\over2}-{11\over 48\pi^2}(gB)^2\ln\left({\Lambda^2\over 
gB}\right),
\label{sv}
\end{equation}
where the classical energy density ${B^2\over 2}$ is weakly screened 
by the zero-point oscillations of the ``gluon'' modes. Instead of 
eq.(\ref{sv}), the variational calculation was found to yield for 
$E(B)$:
\begin{equation}
E(B) = -{11\over48\pi^2}(gB)^2\ln\left({\Lambda^2\over 
gB}\right) + O[(gB)^2],
\label{egp}
\end{equation}
where the classical term ${B^2\over2}$ turns out to be {\it 
completely} screened by the ``unstable oscillators'' of the quantized
gauge-field, characterized by negative squared energies in the 
constant chromomagnetic background, neglected  by Savvidy.

The change from eq.(\ref{sv}) to (\ref{egp}) has dramatic consequences 
for the structure of the GS, for the PGS is seen to lie above the new 
vacuum by the energy density ($B^*$ is the value that minimizes 
$E(B)$),
\begin{equation}
-\Delta E(B^*)= E(0) - E(B^*) ={11\Lambda^4\over96\pi^2}
\ln\left({\Lambda^2\over gB}\right)
\label{de}
\end{equation}
diverging like the fourth power of the ultraviolet cut-off 
$\Lambda$\cite{gp}. But one can argue\cite{gp} that a better, and 
Lorentz-invariant approximation of the GS of QCD is a peculiar state, 
the Chromomagnetic Liquid (CML), consisting of Weiss domains of 
needle-like shape in rapid rotation, inside which there condenses a 
very strong $(gB^*\simeq\Lambda^2)$ chromomagnetic field.

It has been shown\cite{gp1} that the CML solves the problem of colour 
confinement, giving rise to a very good phenomenology of hadrons and 
their interactions.

\section{The variational problem in QG}

Looking at QG as a non-abelian gauge theory, whose gauge group is the 
Poincare group acting on the tangent spaces, it appears natural to try 
and probe
with a similar strategy the stability of the troublesome PGS of QG. This 
alleged GS is, as we know,filled with the zero-point oscillations of 
the  
gravitational field on a flat, Minkowskian background, and is 
corrected in an uncontrollable way by the non-renormalizable 
interactions, stemming from Einstein's action.

In order to define the problem in analogy with QCD we must first find 
a meaningful energy functional for the quantized gravitational field. 
This can be obtained from the ADM\cite{a} approach to QG in terms of 
the ADM mass
\begin{equation}
E_{ADM}={1\over16\pi G}\int_{\partial\Sigma}dS^k \delta^{ij}
(g_{ik,j}-g_{ij,k}),
\label{adm}
\end{equation}
which represents the energy contained in a volume $V$ measured by an 
asymptotic observer, residing in a pseudoeuclidean region. We now set
\begin{equation}
g_{ik} = \eta_{ik} + h_{ik},
\label{g}
\end{equation}
where $\eta_{ik}$ is a background metric and $h_{ik}$ represents the 
field of quantum fluctuations of the metric around the background.

For $\eta_{ik}$ we take the simplest non-trivial solution of 
sourceless Einstein's gravity, the Schwarzschild metric $\eta^{(s)}_{ik}$ of a 
``wormhole'' (WH) of mass $M$ and Schwarzschild radius $2m=2GM$. However no 
fundamental new difficulty arises if one considers the more 
complicated Reissner-Nordstrom and Ker metrics. The energy of space 
around a WH in then given by\cite{letter}
\begin{equation}
E=M+\sum_{n\ge2}\int_\Sigma d^3x (N^{(0)}
H^{(n)}+N_i^{(0)}H^{i(n)}),
\label{adm2}
\end{equation}
where $H^{(n)}$ and $H_i^{(n)}$ are the $n-th$ order expansion in 
$h_{ij}$ of the Hamiltonian and of $-2\pi^{ij}|_j$ ($|$ denotes 
 the covariant derivative with respect to the background field) 
respectively, while $N^{(\circ)}$ is the lapse-function and $N^{(\circ)}$
the shift-vector of the background metric. Our calculation will now 
consist in minimizing over the variational Gaussian wavefunctions:
\begin{equation}
\Psi_\Gamma[h_{ij}] = \exp -{1\over4}\int_{\vec{x} \vec{y}} h_{ij}(\vec 
x)\Gamma^{ijkl}(\vec{x},\vec{y}) h_{kl}(\vec{y}),
\label{wave} 
\end{equation}
$\Gamma^{ijkl}(\vec{x},\vec{y})$ being a variational function, 
the expectation value of the second order Hamiltonian:
\begin{equation}
H^{(2)} = \int d^3xN^{(0)}
H^{(2)}
(h_{ij}(x), -i\hbar \frac{\delta}{\delta h_{ij}(x)}),
\label{opH}
\end{equation}
i.e.
\begin{equation}
\langle H^{(2)}\rangle_\Gamma \equiv {\int [{\cal D}h] \Psi^*[h_{ij}]
H^{(2)}\left[ h_{ij},-i\hbar {\delta \over \delta h_{ij}}\right]\Psi[h_{ij}]
\over \int [{\cal D}h] \Psi^*[h_{ij}]
\left[ h_{ij},-i\hbar {\delta \over \delta 
h_{ij}}\right]\Psi[h_{ij}]}.
\label{e2}
\end{equation}
Before briefly describing the calculation, let me observe that this 
will only have a meaning if it shall be able to justify, 
{\it a posteriori}, the neglect of the practically intractable terms of the 
Hamiltonian for $n\ge 3$.

The class of gaussian functionals must be further restricted by the 
constraint
\begin{equation}
-i{\delta\over\delta\eta_{ij}}\Psi[h]|_j=0,
\label{gau}
\end{equation}
required in ADM-approach\cite{a} by general covariance.

As in the Yang-Mills case the problem can be reduced to finding the 
eigenvalues and the eigenstates of the second order 
wave-operator\cite{letter,article}( $G$ is the Newton constant, $G=l_p^2$ and 
$l_p=10^{-33}$cm is the Planck length)
\begin{equation}
V^{(2)} = -\frac{1}{16\pi G} \eta^{1/2}
(R^{(2)} + \frac{1}{2}h^k_kR^{(1)})=h_{ij}\hat Q^{ijkl}h_{kl},
\label{p}
\end{equation}
where $\eta$ is the determinant of the background metric and 
$R^{(1,2)}$ are the first- and second-order expansion in $h_{ij}$ of 
the scalar of curvature. Setting,
\begin{equation}
\hat{Q}_{ij}^{~~kl}\Phi ^{(\rho)}_{kl} = \lambda(\rho) 
\Phi ^{(\rho)}_{ij},
\label{spettroq}
\end{equation}
where $\Phi ^{(\rho)}_{kl}$ is a complete orthonormal system of second rank 
tensors 
satisfying:
\begin{equation}
\nabla _i (\frac{\Phi ^i_j}{N}) = 0 ~,~ \Phi ^k_k = 0
\label{gaugephi}
\end{equation}
we easily obtain for the minimum energy of the system WH plus quantized 
gravitational field:
\begin{equation}
E = M + {\hbar\over2} \sum_{\rho}\sqrt{\lambda(\rho)}+ O(\hbar^2),
\label{hofame}
\end{equation}
provided, naturally, that $\lambda(\rho)>0$.

\section{The instability of flat space-time}

In the course of a rather difficult and elaborate 
analysis\cite{letter,article} 
aimed at diagonalizing the operator $\hat Q^{ijkl}$ we have made the 
following discoveries:
\begin{itemize}
\begin{enumerate}

\item there exist an $S$-wave mode with negative eigenvalue,
\begin{equation}
\lambda = - {1 \over 64m^2} = - {1 \over 64 (GM)^2}
\label{e}
\end{equation}

\item The stable modes at high energy ($\lambda$ large and positive) in 
the WKB approximation in the Schwarzschild background are red-shifted and 
realize an energy gain with respect to the zero-point energy of the 
gravitational field over flat space-time,
\begin{equation}
\Delta E (M) \simeq - {64 \Lambda^4 R^2 \over \pi^3 } G M {\hbox 
{log}} {R \over 2GM},
\label{dee}
\end{equation}
where $R\gg GM$ is the radius of the spherical volume around the WH, 
and $\Lambda$ the ultraviolet cut-off.
\end{enumerate}
\end{itemize}

It is already clear from the large energy gain (\ref{dee}) that 
flat-space-time is fundamentally unstable, the creation of a WH of mass
$M\ll {R\over G}$ according to (\ref{dee}) leads to a large energy 
advantage that an open, quantized system cannot fail to exploit. But 
there is more. The existence of an unstable mode (see eq.(\ref{e})), 
like in the Yang-Mills case, implies the breakdown of the 
loop-expansion of the energy (see eq.(\ref{dee})) in such a way that the 
negative $O(\hbar)$-oscillation is ``stabilized'' by the higher-order 
($O(\hbar^2)$ and higher) terms that have been neglected. This means 
that the unstable mode will in general produce a {\it negative classical} 
$O(1)$ contribution to the energy. And, just like in the Yang-Mills 
case \cite{prep1,gp}, we expect that its {\it contribution will just cancel 
the classical $M$-term of the theory}. In fact we know that the 
classical ground state of General Relativity (GR) is just flat 
space-time and that its ADM energy is {\it zero}, thus the minimum 
value of the classical energy term of QG is obtained when $M$ is 
precisely compensated by the unstable mode. In order to check how this 
could happen in practice we have compute the Riemann tensor of the 
metric
\begin{equation}
g_{ij} =\eta^{(s)}_{ij} + \mu\Phi_{ij}^{(u)}
\label{gu}
\end{equation}
where $\Phi_{ij}^{(u)}$ 
is the normalized wave-function of the unstable mode and 
$\mu$ is an adjustable amplitude. Averaging the components of the 
Riemann tensor over a spherical shell of width $m=GM$ outside the 
horizon of the WH for a well defined value of $\mu$ we have obtained 
values of its components much smaller than those for $\mu=0$. This 
strongly indicates
that the unstable mode around a WH adjusts its amplitude in such a way 
as to render the space outside the WH horizon {\it as flat as 
possible} and, as a consequence, the ADM energy {\it as small as 
possible}. 

From this semi-quantitative analysis we may conclude that eq.(\ref{dee}) 
should be as adequate approximation of the difference $\Delta E(M,R)$
between the energy of the quantized gravitational field in a spherical 
region of radius R fluctuating around a WH of mass $M\ll {R\over 2G}$
and around flat space-time. Its negative sign shows unequivocally 
that flat space-time, the background of perturbative QG and of the 
Perturbative Ground State, cannot sustain a stable quantum dynamics
of Einstein's General Relativity (GR), thus excluding that the well 
known perturbative pathologies of GR may be of any relevance for QG, 
and of any support for the widely accepted idea that the quantization
of GR is necessarily doomed to fail, requiring a geometry of 
space-time that goes beyond the simple one embodied in GR.

\section{Gas of WH's and the Planck lattice}

We have just seen that the presence of a single WH of mass $M$, in a 
region of radius $R\gg 2MG$, realizes for the quantum gravitational 
field a definite large energy advantage [see eq.(\ref{dee})], with 
respect to empty (classical) Euclidean space, which thus turns out to 
be quantum-mechanically unstable. However, the structure of 
eq.(\ref{dee}) 
makes it quite obvious that the state considered does by no means 
realize the GS of QG, i.e.~the state of {\it minimum} energy, for the 
energy can be further lowered by considering a multiwormhole 
configuration for the classical solution of the sourceless Einstein 
equations. Even though, to my knowledge, we do not know the most 
general multiwormhole solution\footnote{For a particular configuration 
see ref.\cite{mwh}.}, the only important notion that we need in order to 
figure out a good candidate for the GS of QG is that a system of WH's 
of ADM-mass $M$ (and horizon $2GM$) of average density $\left({1\over 
a}\right)^3$ ( $a$ being the average inter WH distance) for $a \geq 4GM$
interacts through a two-body Newtonian potential $V\simeq -{GM^2\over 
a}$. Thus in order to pack a given volume of space $\Omega$ with as 
many elementary spherical regions of radius $\simeq {a\over2}$, each 
realizing the energy gain appearing in eq.(\ref{gu}), it is necessary
that the (semi)-classical system of WH's does not collapse into a 
single giant WH. It is quite easy to find (qualitatively) the 
condition that the stability from collapse puts upon the mass $M$ of 
the WH's. We need only to look at two WH's as (extended) quantum 
particles of mass $M$ and solve the Schr\"odinger equation with reduced 
mass $\mu={M\over2}$ and Newtonian tow-body potential $V=-{GM^2\over 
r}$, whose well known solution yields for the ``Bohr radius''
\begin{equation}
a_\circ={2\over GM^3},
\end{equation}
which represents the average distance between the WH's in the state of 
the lowest energy $E_\circ=2M-{G^2M^5\over4}$. Thus, provided
\begin{equation}
a_\circ\geq 4GM,
\end{equation}
or
\begin{equation}
M\leq G^{-{1\over2}} \left({1\over2}\right)^{1\over4},
\end{equation}
the semi-classical system of WH's will be stable against collapse and 
shall realize the energy density gain with respect to Euclidean 
space-time [$a\simeq 4GM=2^{7\over4}l_p; M\simeq 
m_p\left({1\over2}\right)^{1\over4}, \kappa$ is a number of $O(1)$]
\begin{equation}
{\Delta E\over V}\simeq -\kappa {\Lambda^4\over\pi^3}.
\end{equation}

At this point it should be evident that the semi-classical state over 
which  the quantized gravitational field fluctuates has just the 
features and the structure of the space-time foam, hypothesized many 
years ago by J.A.~Wheeler\cite{w}. It should likewise be evident that 
the observable gravitational field and for that matter all quantum 
fields, that live in the interstices of the fluctuating gas of WH's,
cannot be resolved in space regions of size smaller than the Planck 
length $l_p\simeq 10^{-33}$cm, and should be mathematically described 
by discrete fields defined on a lattice of lattice constant 
$a\simeq l_p$, the Planck lattice.

\section{Conclusion}

In spite of the considerable technical weaponry that has been 
necessary to address a fundamental problem such as determining the 
stability of the PGS of QG, and the structure of its possible GS, it 
is extremely pleasing that the final results of a long and difficult
investigation\cite{letter,article} are extremely simple, and loaded with 
far-reaching consequences and implications.

In a nutshell the main result od this analysis is that Euclidean 
space-time, the classical vacuum, once the metric field is allowed to 
preform quantum fluctuations ( as required by the fundamental 
principles 
of quantum physics) becomes unstable against decaying into a gas of 
WH of mass $M\simeq G^{-{1\over2}}=m_p$ and interwormhole distance
$a\simeq G^{1\over2}=l_p$. This remarkable metamorphosis, that the 
classical vacuum experiences under the ``spell'' of quantum 
fluctuations, has its origin in the simple fact that the 
red-shifts, that the gravitational waves are subject to in the 
surroundings of the WH's, vastly decrease the energy density of the 
zero-point fluctuations upon the classical, Euclidean vacuum. And this 
without paying the price, as in classical physics, of the ADM - 
wormhole mass $M$. Both stunning occurrences had been already 
encountered in a similar analysis of another non-abelian gauge theory, 
the $SU(2)$ Yang-Mills theory\cite{prep1,gp}, that has finally led to a 
simple
and physically transparent proof of colour confinement\cite{gp1}. And it 
is extremely reasonable that similar theoretical structures in the end 
provide us with (most likely) Ground States of similar highly 
non-perturbative character, and this, as far as I can judge, in a 
theoretically robust fashion.

What are the consequences and implications of what has been discovered (and 
summarily described in this talk)? The first and most important 
consequence is that at the Planck scale $l_p$ the familiar continuous 
space-time is found instead to be ``foamy'', i.e.~essentially 
discrete. Apart from realizing another prophecy of Berhhard 
Riemann\cite{re}, this discovery gives us finally the long-sought 
momentum cut-off that is necessary to give a well-defined mathematical 
meaning to all QFT's, that in continuous space-time have been plagued 
with the nasty ultraviolet divergences, that have puzzled great 
minds such as Dirac's. 

The second implication is that the cut-off at the Planck mass gives QG 
also a well defined, perturbative meaning , for the horrible looking, 
formerly non-renormalizable higher order (in the graviton 
fields) interaction terms of the Einstein's action, yield now small 
corrections, calculable in principle. Finally the emergence of the 
``foamy'' space-time, fully justifies a research program into the 
deep structure of the Standard Model\footnote{For review see 
ref.\cite{xuepre}.} that has already produced a number of interesting 
results.

\end{document}